\documentclass{llncs}

\usepackage[T1]{fontenc}
\usepackage[utf8]{inputenc}
\usepackage[english]{babel}

\usepackage{multirow,booktabs}
\usepackage{relsize}

\usepackage[usenames,svgnames]{xcolor}
\usepackage[colorlinks,allcolors=MediumBlue,breaklinks]{hyperref}
\usepackage{graphicx}

\newcommand{\doi}[1]{\href{https://doi.org/#1}{doi:\texttt{#1}}}
\begin{document}

\title{Bibliometric-enhanced Information Retrieval\\ 10th Anniversary Workshop Edition}
\titlerunning{BIR 2020 workshop}  
\author{Guillaume Cabanac\inst{1}\and Ingo Frommholz\inst{2} \and Philipp Mayr\inst{3}}
\authorrunning{Cabanac et al.} 
\institute{
University of Toulouse, Computer Science Department, IRIT UMR 5505, France\\
\email{guillaume.cabanac@univ-tlse3.fr}\\
\and
Institute for Research in Applicable Computing,\\ University of Bedfordshire, Luton, UK,\\\email{ifrommholz@acm.org}
\and
GESIS -- Leibniz-Institute for the Social Sciences, Cologne, Germany,\\
\email{philipp.mayr@gesis.org}%
}

\maketitle

\begin{abstract}	
The Bibliometric-enhanced Information Retrieval workshop series (BIR) was launched at ECIR in 2014 \cite{MayrEtAl2014} and it was held at ECIR each year since then. This year we organize the 10th iteration of BIR. The workshop series at ECIR and JCDL/SIGIR tackles issues related to academic search, at the crossroads between Information Retrieval, Natural Language Processing and Bibliometrics. In this overview paper, we summarize the past workshops, present the workshop topics for 2020 and reflect on some future steps for this workshop series.
\end{abstract}



\begin{table}[ht]\centering
    \caption{Overview of the BIR workshop series}\label{tab:BIR-overview}
    \begin{tabular}{*3{l@{\hspace{.5cm}}}*2{c@{\hspace{.5cm}}}c}\toprule
    	Year & Conference & Venue & Papers & Proceedings \\\midrule
    	2014 & ECIR & Amsterdam,\hfill{} NL & 6   & \href{http://ceur-ws.org/Vol-1143/}{Vol-1143}  \\ 
    	2015 & ECIR & Vienna,\hfill{} AT & 6    & \href{http://ceur-ws.org/Vol-1344/}{Vol-1344}  \\ 
    	2016 & ECIR & Padua,\hfill{} IT & 8   & \href{http://ceur-ws.org/Vol-1567}{Vol-1567} \\ 
    	2016 & JCDL & Newark,\hfill{}US & 10 + 10\textsuperscript{a}   &  \href{http://ceur-ws.org/Vol-1610}{Vol-1610} \\ 
    	2017 & ECIR & Aberdeen,\hfill{}UK & 12   &  \href{http://ceur-ws.org/Vol-1823}{Vol-1823} \\ 
    	2017 & SIGIR & Tokyo,\hfill{}JP & 11   &  \href{http://ceur-ws.org/Vol-1888}{Vol-1888} \\ 
    	2018 & ECIR & Grenoble,\hfill{}FR & 9    & \href{http://ceur-ws.org/Vol-2080}{Vol-2080} \\ 
    	2019 & ECIR & Cologne,\hfill{}DE & 14    & \href{http://ceur-ws.org/Vol-2345}{Vol-2345} \\
    	2019 & SIGIR & Paris,\hfill{}FR & 16 + 10\textsuperscript{b}   & \href{http://ceur-ws.org/Vol-2414}{Vol-2414}  \\ 
    	2020 & ECIR & Lisbon,\hfill{}PT & TBA   & TBA \\\bottomrule\\[-8pt]
    	\multicolumn{5}{l}{\smaller\textsuperscript{a} with CL-SciSumm 2016 Shared Task; \textsuperscript{b} with CL-SciSumm 2019 Shared Task}
    \end{tabular}\\
\end{table}

\section{Motivation and Relevance to ECIR} 
	Searching for scientific information is a long-lived user need.  In the early 1960s, Salton was already striving to enhance information retrieval by including clues inferred from bibliographic citations \cite{Salton1963}.
	The development of citation indexes pioneered by Garfield~\cite{Garfield1955} proved determinant for such a research endeavour at the crossroads between the nascent fields of Bibliometrics\footnote{Bibliometrics refers to the statistical analysis of the academic literature \cite{Pritchard1969} and plays a key role in scientometrics: the quantitative analysis of science and innovation \cite{LeydesdorffAndMilojevic2015}.} and  Information Retrieval (IR)~--- BIR.  The pioneers who established these fields in Information Science~--- such as Salton and Garfield~--- were followed by scientists who specialised in one of these \cite{WhiteAndMcCain1998}, leading to the two loosely connected fields we know of today.

	The purpose of the BIR workshop series founded in 2014 is to tighten up the link between IR and Bibliometrics \cite{MayrAndScharnhorst2015a}.  We strive to get the ‘retrievalists’ and ‘citationists’ \cite{WhiteAndMcCain1998} active in both academia and the industry together, who are developing search engines and recommender systems such as ArnetMiner, Dimensions, Google Scholar, Microsoft Academic Search, and Semantic Scholar, just to name a few.

	These bibliometric-enhanced IR systems must deal with the multifaceted nature of scientific information by searching for or recommending academic papers, patents, venues (i.e., conferences or journals), authors, experts (e.g., peer reviewers), references (to be cited to support an argument), and datasets.  The underlying models harness relevance signals from keywords provided by authors, topics extracted from the full-texts, co-authorship networks, citation networks, and various classifications schemes of science.
  
	BIR is a hot topic with growing recognition in the community in recent years: see for instance the Initiative for Open Citations~\cite{Shotton2018}, the Google Dataset Search~\cite{brickley2019}, the Indian JNU initiative for indexing the world's literature in full-text~\cite{Pulla2019}, the increasing number of retractions~\cite{BrainardAndYou2018}, and massive studies of self-citations~\cite{VanNoorden2019,kacem2019}.  We believe that BIR@ECIR is a much needed scientific event for the ‘retrievalists’, ‘citationists’ and others to meet and join forces pushing the knowledge boundaries of IR applied to literature search and recommendation.

\section{Summarzing the Past BIR Workshops}\label{sec:past}
    The BIR workshop series was launched at ECIR in 2014 \cite{MayrEtAl2014} and it was held at ECIR each year since then.  
    As our workshop lies at the crossroads between IR and NLP, we also ran BIR as a joint workshop called BIRNDL (Bibliometric-enhanced IR and NLP for Digital Libraries) at the JCDL \cite{CabanacEtAl2016} and SIGIR \cite{ChandrasekaranAndMayr2019birndl} conferences. All past workshops had a large number of participants (between $\sim$30 and $\sim$60), demonstrating the relevance of the workshop's topics. 
    
    In the following, we present an overview of the past BIR workshops and keynotes at BIR (Tab.~\ref{tab:BIR-overview}--\ref{tab:BIR-keynote}). All pointers to the workshops and proceedings are hosted at \texttt{\href{https://sites.google.com/view/bir-ws}{sites.google.com/view/bir-ws}}. Many of the presented workshop papers appeared in extended form in one of our four BIR-related special issues (2015 \cite{MayrAndScharnhorst2015a}, 2018 \cite{CabanacEtAl2018a,MayrEtAl2018a}, 2019 \cite{atanassova2019}). 

\section{Workshop Topics}



    The call for papers for the 2020 workshop (the 10th BIR edition) addressed current research issues regarding 3 aspects of the search/recommendation process:
    \begin{enumerate}
        \item User needs and behaviour regarding scientific information, such as:
            \begin{itemize}
                \item Finding relevant papers/authors for a literature review.
                \item Measuring the degree of plagiarism in a paper.
                \item Identifying expert reviewers for a given submission.
                \item Flagging predatory conferences and journals.
                \item Information seeking behaviour and HCI in academic search.
            \end{itemize}        
        \item Mining the scientific literature, such as:
            \begin{itemize}
                \item Information extraction, text mining and parsing of scholarly literature.
                \item Natural language processing (e.g., citation contexts).
                \item Discourse modelling and argument mining.
            \end{itemize}
        \item Academic search/recommendation systems:
            \begin{itemize}
                \item Modelling the multifaceted nature of scientific information.
                \item Building test collections for reproducible BIR.
                \item System support for literature search and recommendation.
            \end{itemize}        
    \end{enumerate}

\begin{table}[ht]\centering
    \renewcommand{\arraystretch}{1.4}
	\caption{Keynotes at BIR}\label{tab:BIR-keynote} %
\begin{tabular}{c@{\hspace{12pt}}l@{\hspace{12pt}}p{7.2cm}@{\hspace{12pt}}l}\toprule
	Year & Area\textsuperscript{a} & Title of the keynote presentation & Presenter\\\midrule
	2015 & SCIM & \raggedright In Praise of Interdisciplinary Research through Scientometrics & Cabanac \cite{DBLP:conf/ecir/Cabanac15}\\ 
	2016 & IR & Bibliometrics in Online Book Discussions:\newline Lessons for Complex Search Tasks & Koolen \cite{DBLP:conf/ecir/Koolen16}\\
	2016 & SCIM & Bibliometrics, Information Retrieval and Natural Language Processing: Natural Synergies to Support Digital Library Research & Wolfram \cite{DBLP:conf/jcdl/Wolfram16}\\
    2017 & IR & Real-World Recommender Systems for Academia: The Pain and Gain in Building, Operating, and Researching them & Beel \cite{DBLP:conf/ecir/BeelD17}\\	
    2017 & NLP & Do “Future Work” sections have a purpose?\newline Citation links and entailment for global \newline scientometric questions & Teufel \cite{DBLP:conf/sigir/Teufel17}\\
    2018 & NLP & Trends in Gaming Indicators: On Failed Attempts at Deception and their Computerised Detection & Labbé \cite{DBLP:conf/ecir/Labbe18}\\	
    2018 & IR & Integrating and Exploiting Public Metadata Sources in a Bibliographic Information System & Schenkel \cite{DBLP:conf/ecir/Schenkel18}\\	
	2019 & NLP & Beyond Metadata: the New Challenges in Mining Scientific Papers & Atanassova \cite{DBLP:conf/ecir/Atanassova19}\\	
	2019 & IR & Personalized Feed/Query-formulation, Predictive Impact, and Ranking & Wade \cite{DBLP:conf/sigir/WadeW19}\\	
	2019 & NLP & Discourse Processing for Text Analysis:\newline Recent Successes, Current Challenges & Webber \cite{DBLP:conf/sigir/Webber19}\\	
	2020 & SCIM & Metrics and trends in assessing the scientific impact & Tsatsaronis\\\bottomrule\\[-18pt]
	\multicolumn{4}{l}{\smaller\textsuperscript{a} SCIM: Scientometrics; NLP: Natural Language Processing; IR: Information Retrieval}
\end{tabular} 
\end{table}

\section{Target Audience}
    The target audience of the BIR workshops are researchers and practitioners, junior and senior, from Scientometrics as well as Information Retrieval and Natural Language Processing. These could be IR/NLP researchers interested in potential new application areas for their work as well as researchers and practitioners working with, for instance, bibliometric data and interested in how IR/NLP methods can make use of such data.

\section{Peer Review Process and Workshop Format} 
    Our peer review process is supported by \href{https://easychair.org}{Easychair}.  Each submission is assigned to 2 to 3 reviewers, preferably at least one expert in IR and one expert in Bibliometrics or NLP. 
	The accepted papers are either long papers (15-minute talks) or short papers (5-minute talks).  Two interactive sessions close the morning and afternoon sessions with posters and demos, allowing attendees to discuss the latest developments in the field and opportunities (e.g., shared tasks such as the CL-SciSumm~\cite{JaidkaEtAl2018} at the BIRNDL joint workshop, see Sect.~\ref{sec:past}).
	These interactive sessions serve as ice-breakers, sparking interesting discussions that usually continue during lunch and the cocktail party. The sessions are also an opportunity for our speakers to further discuss their work.
	
	

\section{Next Steps} 

Research on scholarly document processing has for many years been scattered across multiple venues like ACL, SIGIR, JCDL, CIKM, LREC, NAACL, KDD, and others.
Our next strategic step is the First Workshop on Scholarly Document Processing (SDP)\footnote{\url{https://ornlcda.github.io/SDProc/}} will be held in November 2020 in conjunction with the 2020 Conference on Empirical Methods in Natural Language Processing. This workshop and initiative will be organized by a diverse group of researchers (organizers from BIR, BIRNDL, Workshop on Mining Scientific Publications/WOSP and Big Scholar) which have expertise in NLP, ML, Text Summarization/Mining, Computational Linguistics, Discourse Processing, IR, and others.


\subsubsection{Acknowledgement}
We organizers wish to thank all those who contributed to this workshop series: The researchers who contributed papers, the many reviewers who generously offered their time and expertise, and the participants of the BIR and BIRNDL workshops. 
Since 2016, we maintain the \href{https://github.com/PhilippMayr/Bibliometric-enhanced-IR_Bibliography/}{Bibliometric-enhanced-IR Bibliography} that collects scientific papers which appear in collaboration with the BIR/BIRNDL organizers.

\bibliography{bir-2020}

\begin{thebibliography}{10}
\providecommand{\url}[1]{\texttt{#1}}
\providecommand{\urlprefix}{URL }
\providecommand{\doi}[1]{https://doi.org/#1}

\bibitem{DBLP:conf/ecir/Atanassova19}
Atanassova, I.: Beyond metadata: the new challenges in mining scientific
  papers. In: Cabanac, G., Frommholz, I., Mayr, P. (eds.) Proceedings of the
  8th International Workshop on Bibliometric-enhanced Information Retrieval
  {(BIR} 2019) co-located with the 41st European Conference on Information
  Retrieval {(ECIR} 2019), Cologne, Germany, April 14, 2019. {CEUR} Workshop
  Proceedings, vol.~2345, pp. 8--13. CEUR-WS.org (2019),
  \url{http://ceur-ws.org/Vol-2345/paper1.pdf}

\bibitem{atanassova2019}
Atanassova, I., Bertin, M., Mayr, P.: Editorial: {Mining} {Scientific}
  {Papers}: {NLP}-enhanced {Bibliometrics}. Frontiers in Research Metrics and
  Analytics  (2019). \doi{10.3389/frma.2019.00002}

\bibitem{DBLP:conf/ecir/BeelD17}
Beel, J., Dinesh, S.: Real-world recommender systems for academia: The pain and
  gain in building, operating, and researching them. In: Mayr, P., Frommholz,
  I., Cabanac, G. (eds.) Proceedings of the Fifth Workshop on
  Bibliometric-enhanced Information Retrieval {(BIR)} co-located with the 39th
  European Conference on Information Retrieval {(ECIR} 2017), Aberdeen, UK,
  April 9th, 2017. {CEUR} Workshop Proceedings, vol.~1823, pp. 6--17.
  CEUR-WS.org (2017), \url{http://ceur-ws.org/Vol-1823/paper1.pdf}

\bibitem{BrainardAndYou2018}
Brainard, J., You, J.: What a massive database of retracted papers reveals
  about science publishing’s “death penalty”. Science  (2018).
  \doi{10.1126/science.aav8384}

\bibitem{brickley2019}
Brickley, D., Burgess, M., Noy, N.: Google {Dataset} {Search}: {Building} a
  search engine for datasets in an open {Web} ecosystem. In: The {World} {Wide}
  {Web} {Conference} on - {WWW} '19. pp. 1365--1375. ACM Press (2019).
  \doi{10.1145/3308558.3313685}

\bibitem{DBLP:conf/ecir/Cabanac15}
Cabanac, G.: In praise of interdisciplinary research through scientometrics.
  In: Mayr, P., Frommholz, I., Mutschke, P. (eds.) Proceedings of the Second
  Workshop on Bibliometric-enhanced Information Retrieval co-located with the
  37th European Conference on Information Retrieval {(ECIR} 2015), Vienna,
  Austria, March 29th, 2015. {CEUR} Workshop Proceedings, vol.~1344, pp. 5--13.
  CEUR-WS.org (2015), \url{http://ceur-ws.org/Vol-1344/paper1.pdf}

\bibitem{CabanacEtAl2016}
Cabanac, G., Chandrasekaran, M.K., Frommholz, I., Jaidka, K., Kan, M.Y., Mayr,
  P., Wolfram, D. (eds.): {BIRNDL'16: Proceedings of the Joint Workshop on
  Bibliometric-enhanced Information Retrieval and Natural Language Processing
  for Digital Libraries co-located with the Joint Conference on Digital
  Libraries}, vol.~1610. CEUR-WS, Aachen (2016)

\bibitem{CabanacEtAl2018a}
Cabanac, G., Mayr, P., Frommholz, I.: Bibliometric-enhanced information
  retrieval: {P}reface. Scientometrics  \textbf{116}(2),  1225--1227 (2018).
  \doi{10.1007/s11192-018-2861-0}

\bibitem{ChandrasekaranAndMayr2019birndl}
Chandrasekaran, M.K., Mayr, P. (eds.): {BIRNDL'19: Proceedings of the 4th Joint
  Workshop on Bibliometric-enhanced Information Retrieval and Natural Language
  Processing for Digital Libraries co-located with the Joint Conference on
  Digital Libraries}, vol.~2414. CEUR-WS, Aachen (2019)

\bibitem{DBLP:conf/sigir/2019birndl}
Chandrasekaran, M.K., Mayr, P. (eds.): Proceedings of the 4th Joint Workshop on
  Bibliometric-enhanced Information Retrieval and Natural Language Processing
  for Digital Libraries {(BIRNDL} 2019) co-located with the 42nd International
  {ACM} {SIGIR} Conference on Research and Development in Information Retrieval
  {(SIGIR} 2019), Paris, France, July 25, 2019, {CEUR} Workshop Proceedings,
  vol.~2414. CEUR-WS.org (2019), \url{http://ceur-ws.org/Vol-2414}

\bibitem{Garfield1955}
Garfield, E.: Citation indexes for science: {A} new dimension in documentation
  through association of ideas. Science  \textbf{122}(3159),  108--111 (1955).
  \doi{10.1126/science.122.3159.108}

\bibitem{JaidkaEtAl2018}
Jaidka, K., Chandrasekaran, M.K., Rustagi, S., Kan, M.Y.: Insights from
  {CL-SciSumm} 2016: {T}he faceted scientific document summarization shared
  task. International Journal on Digital Libraries  \textbf{19}(2--3),
  163--171 (2018). \doi{10.1007/s00799-017-0221-y}

\bibitem{kacem2019}
Kacem, A., Flatt, J., Mayr, P.: Tracking self-citations in academic publishing
  (2019), bio$\chi$iv preprint

\bibitem{DBLP:conf/ecir/Koolen16}
Koolen, M.: Bibliometrics in online book discussions: Lessons for complex
  search tasks. In: Mayr, P., Frommholz, I., Cabanac, G. (eds.) Proceedings of
  the Third Workshop on Bibliometric-enhanced Information Retrieval co-located
  with the 38th European Conference on Information Retrieval {(ECIR} 2016),
  Padova, Italy, March 20, 2016. {CEUR} Workshop Proceedings, vol.~1567, pp.
  5--13. CEUR-WS.org (2016), \url{http://ceur-ws.org/Vol-1567/paper1.pdf}

\bibitem{DBLP:conf/ecir/Labbe18}
Labb{\'{e}}, C.: Trends in gaming indicators: On failed attempts at deception
  and their computerised detection. In: Mayr et~al.
  \cite{DBLP:conf/ecir/2018bir}, pp. 6--15,
  \url{http://ceur-ws.org/Vol-2080/paper1.pdf}

\bibitem{LeydesdorffAndMilojevic2015}
Leydesdorff, L., Milojević, S.: Scientometrics. In: Wright, J.D. (ed.)
  {International Encyclopedia of the Social \& Behavioral Sciences}, vol.~21,
  pp. 322--327. Elsevier, 2nd edn. (2015).
  \doi{10.1016/b978-0-08-097086-8.85030-8}

\bibitem{DBLP:conf/ecir/2018bir}
Mayr, P., Frommholz, I., Cabanac, G. (eds.): Proceedings of the 7th
  International Workshop on Bibliometric-enhanced Information Retrieval {(BIR}
  2018) co-located with the 40th European Conference on Information Retrieval
  {(ECIR} 2018), Grenoble, France, March 26, 2018, {CEUR} Workshop Proceedings,
  vol.~2080. CEUR-WS.org (2018), \url{http://ceur-ws.org/Vol-2080}

\bibitem{MayrEtAl2018a}
Mayr, P., Frommholz, I., Cabanac, G., Chandrasekaran, M.K., Jaidka, K., Kan,
  M.Y., Wolfram, D.: Special issue on bibliometric-enhanced information
  retrieval and natural language processing for digital libraries.
  International Journal on Digital Libraries  \textbf{19}(2--3),  107--111
  (2018). \doi{10.1007/s00799-017-0230-x}

\bibitem{MayrEtAl2014}
Mayr, P., Schaer, P., Scharnhorst, A., Larsen, B., Mutschke, P. (eds.): {BIR'16
  Proceedings of the 1st Workshop on Bibliometric-enhanced Information
  Retrieval co-located with the 36th European Conference on Information
  Retrieval}, vol.~1143. CEUR-WS, Aachen (2014)

\bibitem{MayrAndScharnhorst2015a}
Mayr, P., Scharnhorst, A.: Scientometrics and information retrieval: weak-links
  revitalized. Scientometrics  \textbf{102}(3),  2193--2199 (2015).
  \doi{10.1007/s11192-014-1484-3}

\bibitem{Pritchard1969}
Pritchard, A.: Statistical bibliography or bibliometrics? [{D}ocumentation
  notes]. Journal of Documentation  \textbf{25}(4),  348--349 (1969).
  \doi{10.1108/eb026482}

\bibitem{Pulla2019}
Pulla, P.: The plan to mine the world's research papers. Nature  \textbf{571},
  316--318 (2019). \doi{10.1038/d41586-019-02142-1}

\bibitem{Salton1963}
Salton, G.: Associative document retrieval techniques using bibliographic
  information. Journal of the ACM  \textbf{10}(4),  440–457 (1963).
  \doi{10.1145/321186.321188}

\bibitem{DBLP:conf/ecir/Schenkel18}
Schenkel, R.: Integrating and exploiting public metadata sources in a
  bibliographic information system. In: Mayr et~al.
  \cite{DBLP:conf/ecir/2018bir}, pp. 16--21,
  \url{http://ceur-ws.org/Vol-2080/paper2.pdf}

\bibitem{Shotton2018}
Shotton, D.: Funders should mandate open citations. Nature  \textbf{553}(7687),
  ~129 (2018). \doi{10.1038/d41586-018-00104-7}

\bibitem{DBLP:conf/sigir/Teufel17}
Teufel, S.: Do "future work" sections have a purpose? citation links and
  entailment for global scientometric questions. In: Mayr, P., Chandrasekaran,
  M.K., Jaidka, K. (eds.) Proceedings of the 2nd Joint Workshop on
  Bibliometric-enhanced Information Retrieval and Natural Language Processing
  for Digital Libraries {(BIRNDL} 2017) co-located with the 40th International
  {ACM} {SIGIR} Conference on Research and Development in Information Retrieval
  {(SIGIR} 2017), Tokyo, Japan, August 11, 2017. {CEUR} Workshop Proceedings,
  vol.~1888, pp. 7--13. CEUR-WS.org (2017),
  \url{http://ceur-ws.org/Vol-1888/paper1.pdf}

\bibitem{VanNoorden2019}
Van~Noorden, R., Singh~Chawla, D.: Hundreds of extreme self-citing scientists
  revealed in new database. Nature  \textbf{572}(7771),  578--579 (2019).
  \doi{10.1038/d41586-019-02479-7}

\bibitem{DBLP:conf/sigir/WadeW19}
Wade, A.D., Williams, I.: Personalized feed/query-formulation, predictive
  impact, and ranking. In: Chandrasekaran and Mayr
  \cite{DBLP:conf/sigir/2019birndl}, pp.~6--7,
  \url{http://ceur-ws.org/Vol-2414/paper1.pdf}

\bibitem{DBLP:conf/sigir/Webber19}
Webber, B.: Discourse processing for text analysis: Recent successes, current
  challenges. In: Chandrasekaran and Mayr  \cite{DBLP:conf/sigir/2019birndl},
  pp. 8--14, \url{http://ceur-ws.org/Vol-2414/paper2.pdf}

\bibitem{WhiteAndMcCain1998}
White, H.D., McCain, K.W.: Visualizing a discipline: {A}n author co-citation
  analysis of {Information Science}, 1972--1995. Journal of the American
  Society for Information Science  \textbf{49}(4),  327--355 (1998).
  \doi{b57vc7}

\bibitem{DBLP:conf/jcdl/Wolfram16}
Wolfram, D.: Bibliometrics, information retrieval and natural language
  processing: Natural synergies to support digital library research. In:
  Cabanac, G., Chandrasekaran, M.K., Frommholz, I., Jaidka, K., Kan, M., Mayr,
  P., Wolfram, D. (eds.) Proceedings of the Joint Workshop on
  Bibliometric-enhanced Information Retrieval and Natural Language Processing
  for Digital Libraries {(BIRNDL)} co-located with the Joint Conference on
  Digital Libraries 2016 {(JCDL} 2016), Newark, NJ, USA, June 23, 2016. {CEUR}
  Workshop Proceedings, vol.~1610, pp. 6--13. CEUR-WS.org (2016),
  \url{http://ceur-ws.org/Vol-1610/paper1.pdf}

\end{thebibliography}
\end{document}